\documentclass[12pt]{article}
\usepackage{graphicx}

\begin{document}

\title{\textbf{Wide-field telescopes\\ with a Mangin mirror}}

\author{V. Yu. Terebizh\thanks{98409 Nauchny, Crimea, Ukraine;
 \,E-mail:\, \textsf{vterebizh@yandex.ru}}\\
 \small{\textit{Sternberg Astronomical Institute, Moscow, Russia}}}

\date{\footnotesize{}}

\maketitle

\begin{quote}
\small{\textbf{Abstract}~--- Two all-spherical catadioptric optical systems with a Mangin
mirror are described. The design~A (aperture 500~mm, f/2.0) has flat field of view of
$7^\circ$ in diameter; the design~B (aperture 1000~mm, f/1.7) has $10^\circ$ flat field.
Both designs show near-diffraction-limited images. The $D_{80}$ diameter for the design~A
in the integral waveband 0.45-0.85~mcm varies from $1''.3$ on the optical axis up to
$2''.2$ at the edge of the field ($6.2-10.7\,\mu$m); the corresponding range of the
$D_{80}$ diameter for the design~B is $1''.5-1''.9$ ($12.4-16.2\,\mu$m). The designs
include simple glass types, mainly Schott N-BK7 and fused silica. In case of need, better
images could be attained by a choice of other glass, aspherisation of some surfaces, etc.

\medskip

}
\end{quote}

\section*{Introduction}

The spherical meniscus with a reflective back surface is applied in astronomy
since early investigations by W.~Hamilton~[1814], Mangin~[1876], and
Schuppmann~[1899]. One can find a number of such designs in the books by
Maxwell~[1972], Wilson~[1996], Rutten and van~Venrooij~[1999], and in the cited
there papers. The dual purpose of the present research is to substantially
enlarge an angular field of view of a Mangin-based system and a telescope's
aperture, given spherical form of its surfaces.

\begin{figure}[t]   
   \centering
   \includegraphics[width=0.65\textwidth]{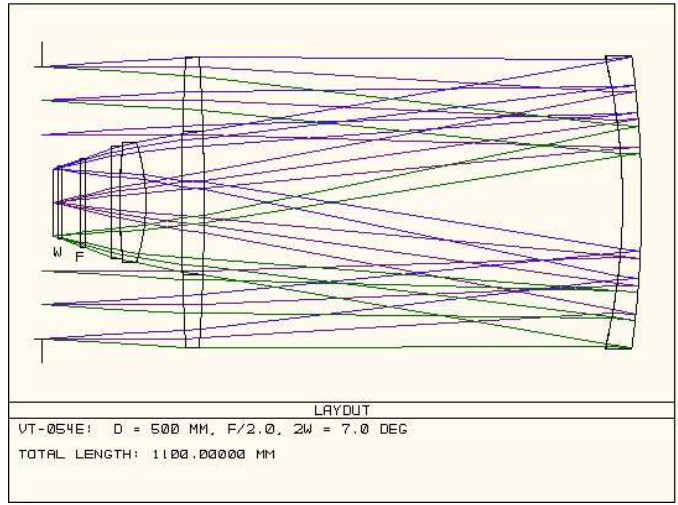}
   \caption*{Design~A. The filter and detector's window \\
     are marked, respectively, by "F" and "W".}
\end{figure}

\section*{Design A}

The layout, the performance, and the complete set of parameters of the design~A are
given, respectively, in Fig.~1, Table~1 of the primary text, and Table~A1 of the
Appendix. The description of surfaces corresponds to the ZEMAX\footnote{ZEMAX Development
Corporation, U.S.A.} optical program.

The basic features of the system A can be described as follows:\\ 1)~The all-spherical
optics, 2)~a remote aperture stop, 3)~a large corrector lens used in the double-pass
mode, 4)~a Mangin mirror, 5)~a two-lens exit corrector made of the simple glass. Let us
discuss the features in brief.

As is well known, the all-spherical optics is not only cheaper at manufacturing, but also
allows to make the surfaces very smooth. Latter property highly promotes increase of the
image quality and mitigates tolerances.

The position of the aperture stop is important both for compensation of aberrations and
for attaining low obscuration of light given such a field size -- we have here only 23\%
of vignetted rays over all field. Even a more important role in providing low vignetting
plays the fact that the large corrector lens works in a double-pass mode. The versions of
the design~A with less field and a hole in the corrector lens were described by
Ceravolo~[2007] and Terebizh~[2007] as the development of the original system of
Hamilton~[1814].

\begin{center}   
\small{
\begin{tabular}{|l|c|c|}
 \multicolumn{3}{l}{\textit{Table 1. Performance of the systems}}\\[3pt]
 \hline
 &&\\
 \multicolumn{1}{|c|}{Parameter} & Design A & Design B \\
 &&\\
 \hline
 &&\\
 Entrance pupil diameter         & 500 mm                & 1000 mm \\
 Effective diameter              & 438 mm                & 853 mm \\
 Effective focal length          & 1000.0 mm             & 1724.1 mm \\
 Effective f-number              & 2.0                   & 1.72 \\
 Scale in the focal plane        & 4.848 $\mu$m/arcsec   & 8.359 $\mu$m/arcsec\\
 Angular field of view           & $7^\circ.0$           & $10^\circ.0$\\
 Linear field of view            & 123.0 mm              & 302.4 mm \\
 Image RMS-diameter              &                       & \\
  \quad center of field          & $0''.68$ (3.3 $\mu$m) & $0''.97$ (8.1 $\mu$m)\\
  \quad edge                     & $1''.5$ (7.2 $\mu$m)  & $1''.4$ (11.7 $\mu$m)\\
 Image $D_{80}$ diameter         &                       & \\
  \quad center of field          & $1''.3$ (6.2 $\mu$m)  & $1''.5$ (12.4 $\mu$m)\\
  \quad edge                     & $2''.2$ (10.7 $\mu$m) & $1''.9$ (16.2 $\mu$m)\\
 Maximum distortion              & 0.50\%                & 0.23\%\\
 Fraction of unvignetted rays    &                       & \\
 \quad center of field -- edge   & 0.769 -- 0.768        & 0.728 -- 0.728 \\
 Length of the optical system    & 974 mm                & 1704 mm \\
 \hline
\end{tabular}
 }
\end{center}

\medskip

Aberrations of a Mangin mirror were repeatedly discussed in literature (see the above
mentioned references), so we only note here that apprehensions connected with more severe
tolerances of an internal mirror surface seem to be groundless. A version of the system~A
of slightly less size has been made recently in Sternberg Astronomical Institute, it
shows the perfect images.

As to the glass types, the basic Schott glass N-BK7 has been used for the large optical
elements, although fused silica is also allowable. There are a few attractive choices of
glasses for two lenses of the exit corrector; we have preferred a pair N-BAK2 plus
N-BASF64 to somewhat better combinations because of their presence in a renewed Schott
catalogue and the low cost, only 2.0 and 3.0 relatively to N-BK7.

Let us consider the image quality of the design~A. Fig.~2 depicts the corresponding spot
diagrams. Since a point-like object produces images of $3.2-4.1\,\mu$m in diameter, when
the wavelength varies from $0.65\,\mu$m up to $0.85\,\mu$m, one can see from Fig.~2 that
the telescope provides nearly diffraction-limited images in the integral light.
Naturally, using of the filters leads to better images.

\begin{figure}[t]   
   \centering
   \includegraphics[width=0.65\textwidth]{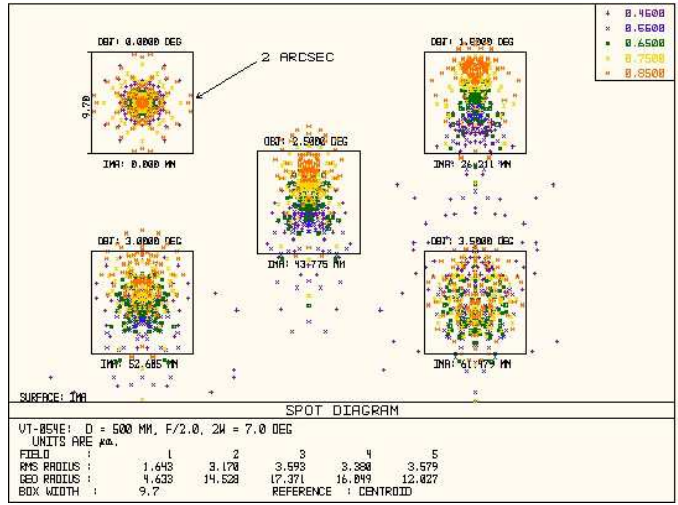}
   \caption*{Spot diagrams for the system~A \\ in the integral light $0.45-0.85\,\mu$m.
    The field angles are \\ $0,\, 1^\circ.5,\, 2^\circ.5,\, 3^\circ.0$, and $3^\circ.5$.
    The box width is $2''$ ($9.7\,\mu$m).}
\end{figure}

\begin{figure}[t]   
   \centering
   \includegraphics[width=0.65\textwidth]{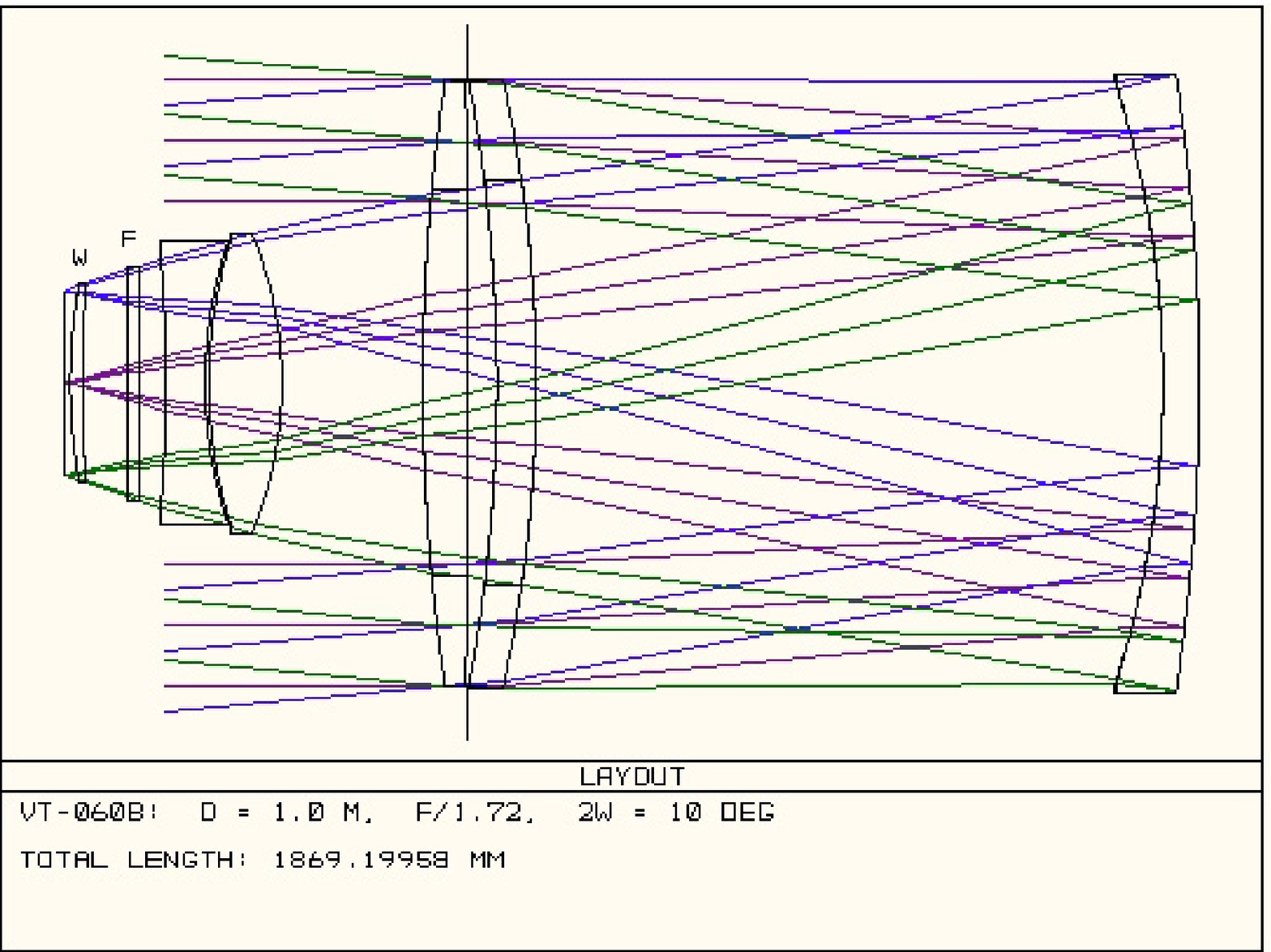}
   \caption*{Design~B. The filter and detector's window \\
    are marked respectively by "F" and "W".}
\end{figure}

\begin{figure}[t]   
   \centering
   \includegraphics[width=0.65\textwidth]{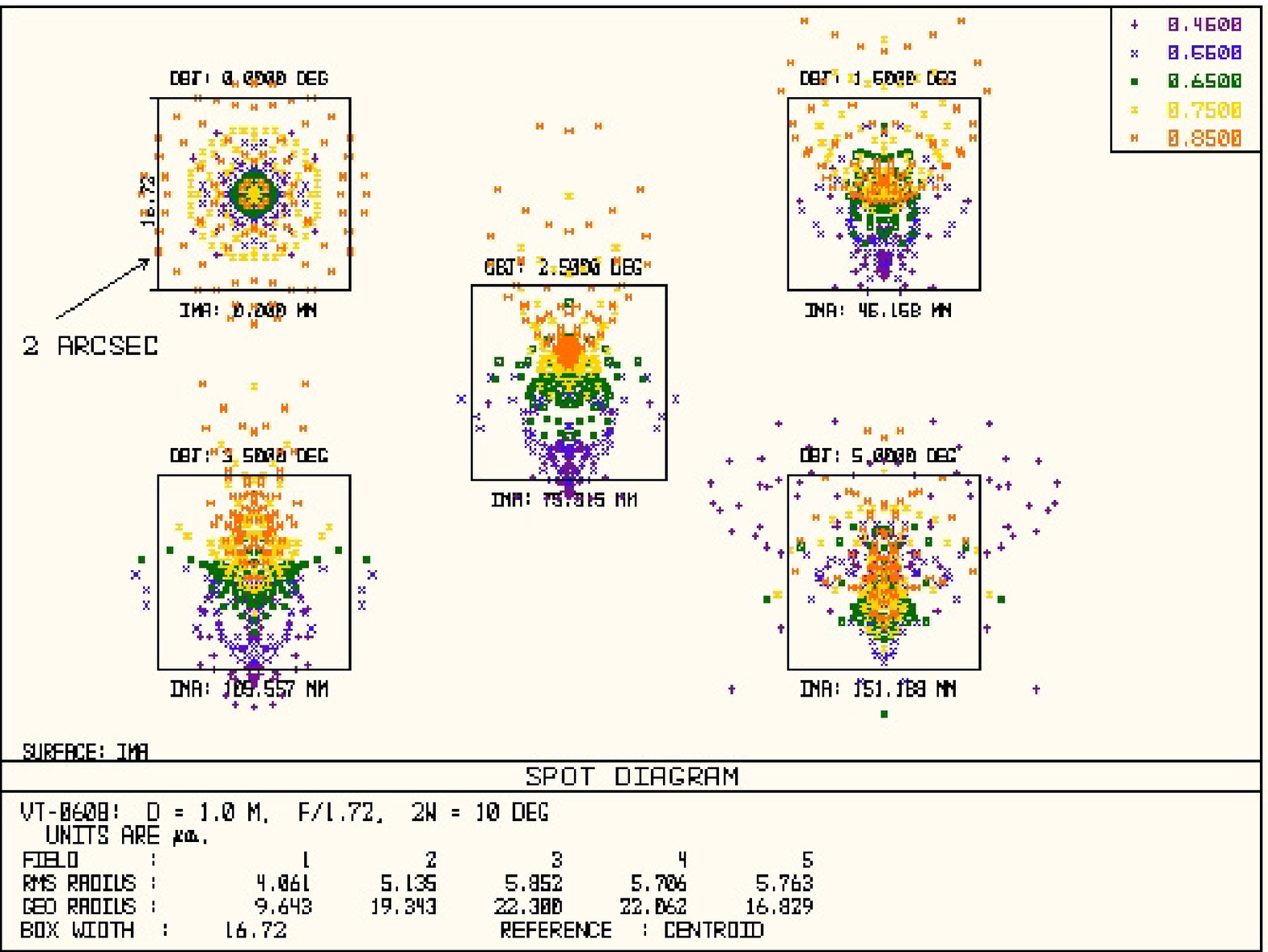}
   \caption*{Spot diagrams for the system~B\\ in the integral light $0.45-0.85\,\mu$m.
    The field angles are \\ $0,\, 1^\circ.5,\, 2^\circ.5,\, 3^\circ.5$, and $5^\circ.0$.
    The box width is $2''$ ($16.7\,\mu$m).}
\end{figure}

\section*{Design B}

Main difference of the design~B from the considered above system consists in an
additional large lens, which is also applied in a double-pass mode (Fig.~3, Tables 1 and
A2). The double-lens entrance corrector has allowed us to reach twice larger aperture and
$10^\circ$-field at nearly the same image quality as before.

The changes of secondary importance are the position of the aperture stop and the curved
detector's window. The stop remains far from the Mangin mirror to efficiently suppress
its aberrations. Since linear size of the field of view is now more than 300~mm, there is
good reason to believe that it is easier to make a slightly curved window lens than a
strictly flat plate.

This time we prefer to use fused silica, instead of N-BK7, for the large optical
elements. The latter glass is applied for the second lens of the exit corrector, while
Schott N-FK5 was chosen for the first lens. There are a number of pairs of glasses for
the exit lens corrector ensuring the somewhat better image quality; N-FK5 and N-BK7 were
chosen because these glasses are available now in the enough large blocks and are cheap
(the relative cost of N-FK5 is 5.0).

\section*{Concluding remarks}

Both considered above systems represent base models intended to attain images about
one-second-of-arc quality in a flat field up to $10^\circ$ in diameter, given spherical
optics and low obscuration of light. If the better images are necessary, one could apply
further known means, in particular, aspherisation of some surfaces and sophisticated exit
corrector. Since the basic designs deliberately were optimized for the spherical set of
surfaces, the first of the mentioned ways is rather inefficient, while the second one is
quite productive. For example, the use of a fluoro-crown (like N-FK51A) or phosphate
crown (like N-PK52A) in a 3- or 4-lens exit corrector provides really sub-arcsecond
images in a wide, as before, field.

It is worth to note also that both systems under consideration are well protected from
stray light.

\medskip

The author is grateful to M.R.~Ackermann and V.V.~Biryukov for useful
discussions.

\section*{Appendix:\\ The complete description of the designs}

\begin{center}   
\begin{tabular}{|c|c|c|c|c|c|}
 \multicolumn{6}{l}{\textit{Table A1. Design~A}}\\[3pt]
\hline
 &&&&&\\
 Number  &         &Curvature &Thickness &      &Light     \\
 of the  &Comments &radius    &(mm)      &Glass &diameter  \\
 surface &         &(mm)      &          &      &(mm)      \\
 &&&&&\\
\hline
 &&&&&\\
 1  &Aperture stop     &$\infty$    &126.34     &---        &500.0 \\
 2  &Shield$^{a}$      &$\infty$    &129.15     &---        &120.0 \\
 3  & Lens 1           &$4803.36$   &41.34      &N-BK7      &532.2 \\
 4  &                  &$-3999.98$  &768.16     &---        &533.3 \\
 5  & Mangin           &$-1156.39$  &35.0       &N-BK7      &531.3 \\
 6  &Primary           &$-2013.58$  &$-35.0$    &Mirror     &538.8 \\
 7  &                  &$-1156.39$  &$-768.16$  &---        &520.8 \\
 8  &Lens 1            &$-3999.98$  &$-41.34$   &N-BK7      &264.7 \\
 9  &                  &$4803.36$   &$-65.73$   &---        &255.5 \\
 10 &Lens 2            &$-351.86$   &$-47.89$   &N-BAK2     &220.7 \\
 11 &                  &$955.57$    &$-0.383$   &---        &208.5 \\
 12 &Lens 3            &$931.71$    &$-15.15$   &N-BASF64   &208.3 \\
 13 &                  &$28228.1$   &$-47.53$   &---        &199.6 \\
 14 &Filter            &$\infty$    &$-10.0$    &N-BK7      &163.6 \\
 15 &                  &$\infty$    &$-34.03$   &---        &158.8 \\
 16 &Window            &$\infty$    &$-5.0$     &F\_SILICA  &133.1 \\
 17 &                  &$\infty$    &$-10.0$    &---        &130.6 \\
 18 &Image             &$\infty$    &           &           &123.0 \\
\hline
\end{tabular}
\end{center}
 {\footnotesize
 $^{a)}$ Circular obscuration between radiuses 0.0 and 120.0~mm.\\
 }

\newpage

\begin{center}   
\begin{tabular}{|c|c|c|c|c|c|}
 \multicolumn{6}{l}{\textit{Table A2. Design~B}}\\[3pt]
\hline
 &&&&&\\
 Number  &         &Curvature &Thickness &      &Light     \\
 of the  &Comments &radius    &(mm)      &Glass &diameter  \\
 surface &         &(mm)      &          &      &(mm)      \\
 &&&&&\\
\hline
 &&&&&\\
 1  &Shield$^{a}$      &$\infty$    &424.08     &---        &520.0  \\
 2  & Lens 1           &$3455.65$    &75.0       &F\_SILICA  &1002.8 \\
 3  &                  &$-28574.4$  &0.25       &---        &998.2 \\
 4  &Aperture stop     &$\infty$    &46.58      &---        &998.4 \\
 5  &Lens 2            &$-2719.94$  &$65.0$     &F\_SILICA  &998.4 \\
 6  &                  &$-2331.60$  &$1032.82$  &---        &1005.9 \\
 7  & Mangin           &$-1611.04$  &60.0       &F\_SILICA  &1000.6 \\
 8  &Primary           &$-3372.22$  &$-60.0$    &Mirror     &1020.5 \\
 9  &                  &$-1611.04$  &$-1032.82$ &---        &980.0 \\
 10 &Lens 2            &$-2331.60$  &$-65.0$    &F\_SILICA  &667.6 \\
 11 &                  &$-2719.94$  &$-46.58$   &---        &648.9 \\
 12 & Stop position    &$\infty$    &$-0.25$    &---        &639.7 \\
 13 &Lens 1            &$-28574.4$  &$-75.0$    &F\_SILICA  &639.0 \\
 14 &                  &$3455.65$   &$-232.56$  &---        &625.1 \\
 15 &Lens 3            &$-646.65$   &$-119.97$  &N-FK5      &495.0 \\
 16 &                  &$793.38$    &$-6.56$    &---        &475.7 \\
 17 &Lens 4            &$740.69$    &$-65.0$    &N-BK7      &467.8 \\
 18 &                  &$-2653.12$  &$-42.41$   &---        &410.6 \\
 19 &Filter            &$\infty$    &$-20.0$    &N-BK7      &385.7 \\
 20 &                  &$\infty$    &$-75.06$   &---        &376.6 \\
 21 &Window            &$2176.5$    &$-20.0$    &N-BK7      &326.8 \\
 22 &                  &$901.89$    &$-8.0$     &---        &322.2 \\
 23 &Image             &$\infty$    &           &           &302.5 \\
\hline
\end{tabular}
\end{center}
 {\footnotesize
 $^{a)}$ Circular obscuration between radiuses 0.0 and 260.0~mm.\\
 }


\begin{thebibliography}{}

\bibitem{} P.~Ceravolo, 2007. See http://www.ceravolo.com/

\bibitem{} W.F.~Hamilton, 1814. Engl. Pat. No. 3781.

\bibitem{} A.~Mangin, 1876. \textit{Memorial de l'officier du genue},
 \textbf{25.(2)10}, 211-289.

\bibitem{} J.~Maxwell, 1972. \textit{Catadioptric Imaging Systems},
 American Elsevier, New York.

\bibitem{} H.G.J.~Rutten, M.A.M.~van~Venrooij, 1999.
 \textit{Telescope Optics}, Willmann-Bell, Richmond.

\bibitem{} L.~Schuppmann, 1899. \textit{Die Medial-Fernrohre},
 B.G. Teubner, Leipzig.

\bibitem{} V.Yu.~Terebizh, 2007. \textit{Wide-field Telescopes}.
 In: \textit{Astronomy: Traditions, Present and Future},
 St.-Petersburg State University, pp. 362-395.

\bibitem{} R.N.~Wilson, 1996. \textit{Reflective Telescope Optics},
 \textbf{I}, Springer.

\end{thebibliography}
\end{document}